# Bond valence calculation for several perovskites and the evidences for a valence charge transfer process in these compounds

Hoang Nam Nhat

Faculty of Physics, Hanoi University of Sciences, VNU,
334 Nguyen Trai, Hanoi, Vietnam

E-mail address: namnhat@hn.vnn.vn (Hoang Nam Nhat)

**Abstract**

This paper presents the bond valence calculation for several perovskite systems and describes the evidences for a valence charge transfer process in these compounds. The reviewing of the crystal structures of $La_{1-x}Pb_xMnO_3$ (x=0.1-0.5), $La_{0.6}Sr_{0.4-x}Ti_xMnO_3$ (x=0.0-0.25) and $La_{1-x}Sr_xCoO_3$ (x=0.1-0.5) is also presented. On the basis of testing samples, the distribution of valence charge has been evaluated which showed the failure of elastic bonding mechanism on all studied systems and revealed the general deficit of valence charge in the unit cell. This deficit was not equally localized on all coordination spheres but proved asymmetrically distributed between the spheres. As the content of substitution increased, the charge deficit declined systematically from balanced level, signifying the continuous transfer of valence charge from the $B-O_6$ to $A-O_{12}$ spheres. The transfered charge varied from system to system, depending on the valence deviation of spheres and was not small. The total valence deviation reached near 2 electron/unit cell in the studied systems. The local deviation may be more larger than this average value. The possible impact of the limitted accuracy of the available structural data on the bond valence results has been considered.

## 1 Introduction

Many application significant properties of perovskites $ABO_3$ are believed to have origin in their structures. Although simple in general, the perovskite structures still pose many questions and among the big ones is the stoichiometry of elements. The common consensus is to consider a small diversity δ from the perfect formula stoichiometry $ABO_3$ i.e. to consider the real formulas as of $ABO_{3-δ}$ or $A_{1-δ}B_{1-δ}O_3$ [25]. Certainly, this consideration of non-stoichiometry forces a different view on valence charge distribution in unit cell. This small amount δ may be determined by various methods, e.g. the chemical titration where applicable or the non-destructive Rietveld analysis of powder diffractions. The later one is not frequently as acurate as required since the Rietveld method suffers a lot from low resolution of powder diffractions. For many perovskites this low resolution usually causes the situation that $R_{Profile}$ factor - the measure of Rietveld method accuracy, develops slowly during refinement process and does not follow radical changes in the structural parameters of samples. It may quite happen that the solutions with better $R_{Profile}$ are physically less meaningful than the worse ones, so the obtained results are principally ambiguous. Either by Rietveld or chemical method, the achieved non-stoichiometry δ always associates with physical structure of materials, that is one has to consider the defects even in the perfect crystal. In this paper we introduce a bond valence approach to the quantitative analysis of perovskite structures and show that these structures may be regarded as having "charge defect" without specifying any physical defect. The study reveals asymmetric charge distribution and certain charge transfer process between the coordination spheres of the metals.

The valence theory of chemical bond originated from 1929 when L. Pauling postulated the so-called *valence principle* [2,3] which says that the atomic valence of any atom X is equal to the sum of valences of all bonds to X: $v = \Sigma v_i$. Evidently, this atomic valence is equivalent to the absolute value of oxidation state of the given atom (e.g. atom $O^{2-}$ has atomic valence of $v = |-2| = 2$). Similarly, the bond valence (BV) is equivalent to the number of bonding electrons (BE) distributed within the bond, for examples the average BE in the coordination sphere $[Co^{2+}O^{2-}_6]$ is 2/6 = 0.333 $e^-$ per bond. The dependence of BV on bond length (BL) was subject of extensive studies for decades [1,4-5,9,28-31] and many functional dependences were published, most of them was summarized in [4]. All these functions are the exponential functions of general form $v_i = e^{(R_0-R)/B}$ where $R_0$ and $B$ are empirical parameters determined particularly for each donor-acceptor pair, usually for the metal-oxygen bonds (**Fig.1**). According to these relationships, when BLs increase BVs must decrease and *vice versa*. This characteristic dependence was postulated by Brown [1] in a principle called the *distortion theorem* which says: "*The product of BV and average BL is constant for the same bonds*" (i.e. $v_i \times \langle R_i \rangle = const$). Although this principle is not valid in general, it proves correct in many cases, especially for the ionic compounds and the hydrogen bonding systems. It is clear that this principle illustrates the elastic bonding mechanism. When BLs vary, BVs must change also to preserve the electric neutrality of molecule and the stoichiometry of total atomic valence.

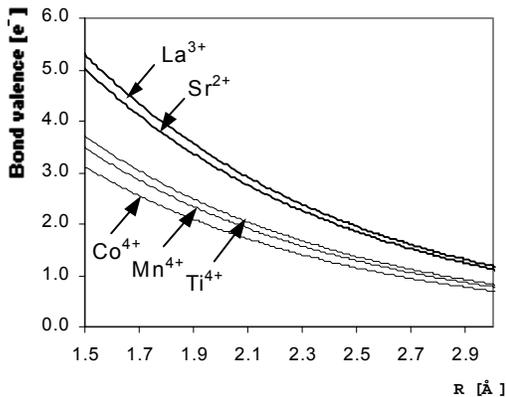

**Fig. 1**. *The bond valence curves for several cations*

The distortion theorem can be used, as we show here, to analyse the bonding properties of perovskites where the substitution of various elements (e.g. $A=La^{3+}$, $Sr^{2+}$, ...; $B=Mn^{3+}$, $Co^{3+}$...) of different oxidation states to the same lattice positions deforms the original 'substitution-free' cubic f-b-c lattice ($A^{3+}B^{3+}O_3^{2-}$). For the perovskites having atom X at lower oxidation state (e.g. $Sr^{2+}$) replacing atom A at higher oxidation state (e.g. $La^{3+}$) with $x\%$, the common opinion is that some portion of atom B (e.g. $Co^{3+}$) must also change its oxidation state to higher (e.g. $Co^{4+}$). This consequently, as stated by distortion theorem, leads to the increase in valences of bonds around atom B and to the shortening of lattice constants connecting atoms B. The experimental results, however as discussed later, did not confirm the contract of lattices. For all studied perovskites, the lattice parameters changed only a little, reflecting no significant variation in bond valences, i.e. the number of electrons distributed within the unit cell remains unchanged. Thus the unit cell must be under-charged comparing to the state predicted by distortion theorem. This situation either forces variation in stoichiometry of some atoms, namely of the oxygens if distortion theorem holds, or signifies the irregular charge relocation phenomenon within perovskite lattice if distortion theorem is considered as failed.

The bond valence method, however, requires the precise determination of interatomic distances and consequently of atomic positions, which is not always possible and available in the literature. To enable the use of this method for a larger group of perovskites we have developed a simple procedure for obtaining the bond valences directly from lattice parameters. In this paper we (a) review the crystal structures of the following perovskites: $La_{1-x}Pb_xMnO_3$ ($x$=0.1-0.5) and $La_{0.6}Sr_{0.4-x}Ti_xMnO_3$ ($x$=0.0-0.25) determined by Rietveld method. It is worth to note here that the re-determination of these two perovskite systems only aimed at providing the additional structural data to ensure the consistency with the data already available, that is they do not differ each from other too much. It is important to understand that the perfect accuracy of structure data is not required (and in many cases impossible to achieve by the powder diffraction method, which mainly applies for the perovskites); (b) present the theoretical concepts for obtaining bond valences and for estimation of non-stoichiometry parameter $\delta$; (c) investigate the bond valence charge distribution and stoichiometry of the above two perovskite systems, of $La_{1-x}Sr_xCoO_3$ (x=0.0-0.5) and of several other systems, drawing attention to the failure of distortion theorem and finally (d) discuss the evidences for a charge transfer process between the coordination spheres of metals in perovskites and point to some possible consequences. Theoretical concepts are given in Section 2 and results are discussed in Section 3. Detailed discussion on failure of distorsion theorem and valence charge transfer phenomenon is found in the Sections 4 and 5. Section 6 provides some remarks and conclusions are given in Section 7.

All samples have been prepared by the conventional solid-state reaction method. For $La_{1-x}Pb_xMnO_3$: the raw powders $La_2O_3$, PbO and $MnCO_3$ were presintered in air at $900^0$C-$1000^0$C for 15 hours then removed, reground and pressed into pellets and sintered at the same temperature for 15hours. For $La_{0.6}Sr_{0.4-x}Ti_xMnO_3$: the raw powders $La_2O_3$, $SrCO_3$, $TiO_2$, $MnCO_3$ were presintered in air at $1250^0$C for 10 hours and then sintered at $1300^0$C for 12 hours. The powder samples were measured at room temperature using the Bruker's X-Ray Diffractometer D5005 at Center for Materials

Science, Faculty of Physics, Hanoi University of Science. Step angle 0.03, from $10^0$ to $70^0$, profile points 2000. Profiles were optimised using Pseudo-Voight functions. Lattice structures were determined by three methods: Ito [6], Visser [7] and Taupin [8]. Rietveld refinement was carried out using program MPROF [5].

## 2 Bond valence theory of perovskites

### 2.1. Perovskites in the 'pseudo-cubic' lattice

The oblique perovskite lattice is the cubic f-b-c with lattice constant $a \sim 3.8$ Å where atom A occupies origin A(0,0,0); B occupies b-c position B(½,½,½); and O occupies 3 f-c positions $O_1$(½,½,0), $O_2$(½,0,½), $O_3$(0,½,½). These 3 f-c positions are equivalent only in the cubic lattice, not in lower symmetry. The coordination number of A is 12 A–$O_{12}$, of B is 6 B–$O_6$, of O is 2+4 (O–$B_2A_4$) (**Fig.2**). The cubic lattice may be deformed to the lower symmetries, e.g. to the monoclinic ($a=c\neq b\sim 3.8$Å, $\alpha=\gamma=90°$, $\beta\sim 90°$), or to the rhombohedral ($a=b=c\sim 3.8$Å, $\alpha=\beta=\gamma\neq 90°$; note that the frequently reported hexagonal lattice $a=b\sim 5.5$Å and $c\sim 6.6$Å or $c\sim 13.0$Å, $\alpha=\beta=90°$, $\gamma=120°$ is equivalent to the rhombohedral one) or even to the triclinic lattice ($a\neq b\neq c\sim 3.8$Å, $\alpha\neq\beta\neq\gamma\sim 90^0$). Some other lattice types, such as tetragonal (a=b=c$\sim$7.6Å), also exist.

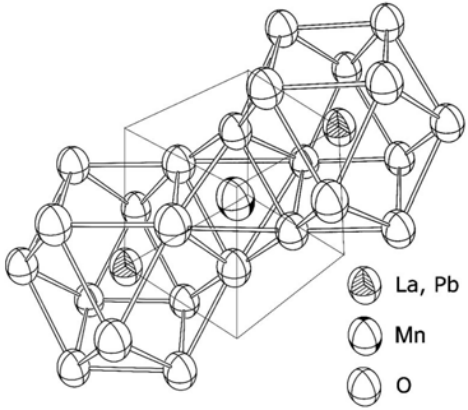

**Fig. 2** *Stacking of coordination polyheders for $(La_{1-x}Pb_x)MnO_3$. The lines connecting the oxygens do not mean the bonds and are drawn only to shape the coordination polyeders.*

Despite the variety of lattices at lower symmetries we have observed for the studied perovskites that in the lattices with $z=1$ (A at the origin [0,0,0] and $a, b, c, \alpha, \beta, \gamma$ near 3.8Å and $90^0$) there was usually no recognizable improvement of $R_{Profile}$ when the atomic positions were refined. For most cases, the Rietveld analysis yielded the best $R_{Profile}$ in P1 space group with all atoms set fixed at the cubic f-b-c positions. This means that the atoms may be considered as resided unchangeably at the elements of f-b-c cubic symmetry, i.e. A at (0,0,0), B at (½,½,½); and O at (½,½,0), (½,0,½), (0,½,½) although the lower symmetries may not possess the elements of symmetry at these positions. This specific behaviour suggested us to call these lattices the *pseudo-cubic* ones. Hereinafter the term pseudo-cubic refers to the distorted cubic lattices at various lower symmetries, e.g. orthorhombic, rhombohedral and triclinic. A very interest feature of this lattice is that it is very *sensitive to angular deformation*, so is suitable for studying the slight symmetry-breaking distortions, which are usually unobserved in the other lattice types. Furthermore, in the pseudo-cubic lattice the atomic distances, bond lengths and bond angles are deducible directly from lattice constants. So for cases where diffraction data do not allow accurate Rietveld analysis of atomic positions we refined the lattices in pseudo-cubic form and derive the bonding parameters directly from lattice constants.

### 2.2. Calculation of BV for pseudo-cubic cell

In general triclinic symmetry P $\bar{1}$ the coordination sphere A–$O_{12}$ of atom A consists from 3 tetragons A$\cdots O_4$; bonds within each tetragon are divided into 2 groups of equal length A–O, so the A–$O_{12}$ polyheder setup up 6 different bond lengths. The coordination sphere B–$O_6$ consists from 3 pairs B$\cdots O_2$ setting up 3 different bond lengths. For oxygen, the coordination sphere O–$B_2A_4$ consists from 1 pair O–$B_2$ plus 2 pairs O–$A_2$ setting up 3 different bond lengths for each oxygen positions $O_1$(½,½,0), $O_2$(½,0,½), $O_3$(0,½,½). Where the interatomic distances are not available, or available but at low accuracy, the bond valences may be derived from the lattice constants as followed. Let $\vec{a}, \vec{b}, \vec{c}$ be the lattice vectors. By using Pauling's bond-valence sum rule $v = \sum v_i = \sum e^{(R_o-R)/B}$ we obtain the following relations for triclinic symmetry:

(1) *Valence of A:*  $v_A = v_{A1} + v_{A2} + v_{A3}$

$$v_{A1} = 2e^{(R_0-|(\vec{a}+\vec{b})/2|)/B} + 2e^{(R_0-|(-\vec{a}+\vec{b})/2|)/B}$$

$$v_{A2} = 2e^{(R_0-|(\vec{a}+\vec{c})/2|)/B} + 2e^{(R_0-|(-\vec{a}+\vec{c})/2|)/B}$$

$$v_{A3} = 2e^{(R_0-|(\vec{c}+\vec{b})/2|)/B} + 2e^{(R_0-|(-\vec{c}+\vec{b})/2|)/B}$$

(2) *Valence of B:*  $v_B = v_{B1} + v_{B2} + v_{B3}$

$$v_{B1} = 2e^{(R_0-|\vec{a}/2|)/B}; v_{B2} = 2e^{(R_0-|\vec{b}/2|)/B}; v_{B3} = 2e^{(R_0-|\vec{c}/2|)/B}$$

(3) *Valence of O:* There are 3 independent positions $O_1$, $O_2$ and $O_3$ so the valence is calculated for each case separately:

$v_{O1} = v_{B-O1} + v_{A-O1}$; $v_{O2} = v_{B-O2} + v_{A-O2}$; $v_{O3} = v_{B-O3} + v_{A-O3}$

$$v_{B-O_1} = 2e^{(R_0-|\vec{a}/2|)/B}; v_{B-O_2} = 2e^{(R_0-|\vec{b}/2|)/B}; v_{B-O_3} = 2e^{(R_0-|\vec{c}/2|)/B}$$

$$v_{A-O_1} = 2e^{(R_0-|(\vec{b}+\vec{c})/2|)/B} + 2e^{(R_0-|(-\vec{b}+\vec{c})/2|)/B}$$

$$v_{A-O_2} = 2e^{(R_0-|(\vec{a}+\vec{c})/2|)/B} + 2e^{(R_0-|(-\vec{a}+\vec{c})/2|)/B}$$

$$v_{A-O_3} = 2e^{(R_0-|(\vec{b}+\vec{a})/2|)/B} + 2e^{(R_0-|(-\vec{b}+\vec{a})/2|)/B}$$

The average valence for atom O is:
$\langle v_O \rangle = (v_{O1} + v_{O2} + v_{O3}) / 3$

(4) *The electric neutrality of molecule requires*:
$v_A + v_B - v_{O1} - v_{O2} - v_{O3} = 0$

This relation only means $v_A + v_B = v_{O1} + v_{O2} + v_{O3}$. It does not say $v_A + v_B = 3 \times |-2| = 6$. In fact this sum may differ from 6 and this diversity signifies the real charge in the unit cell. The non-stoichiometry parameter δ is determined on the basis of this diversity.

Where there is an atom $X^{2+}$ replacing atom $A^{3+}$ with $x\%$ content, the valence spheres $A^{3+}-O_{12}$ change to $A^{3+}-O_{12}$ plus $X^{2+}-O_{12}$ and the spheres $O-A^{3+}_4$ change to $O-A^{3+}_4$ plus $O-X^{2+}_4$. Here we assume that the atoms X replace A homogeneously, that is all oxygen positions $O_1$, $O_2$ and $O_3$ contribute to the change of oxygen valence equally. Let define the average valence $<v_A>$ in the A-position as:

$<v_A> = v_A \times (1-x) + v_X \times x$, where $v_X$ is calculated as for $v_A$ but with $R_0$ for $X^{2+}-O$.

Evidently $<v_A>$ decreases with increasing $x$ since the valence of X (2+) is lower than of A (3+). Therefore, the atoms B tend to compensate this charge deficiency by exhibiting both oxidation states 3+ and 4+. A portion of $B^{3+}$ will move to $B^{4+}$. Suggest that this portion is equal to $x\%$, the average valence at B positions is defined as followed:

$<v_B> = v_{B+3} \times (1-x) + v_{B+4} \times x$, where $v_{B+4}$ is calculated as for $v_{B+3}$ but with $R_0$ for $B^{4+}-O$.

Similarly, for the calculation of oxygen valence $v_O = (v_{O1} + v_{O2} + v_{O3})/3$, the valences $v_{A-O1}$, $v_{A-O2}$, $v_{A-O3}$ at $O_1$, $O_2$ and $O_3$ positions should be replaced by the averages $<v_{A-O1}>$, $<v_{A-O2}>$, $<v_{A-O3}>$ defined as:

$<v_{A-O(i)}> = v_{A-O(i)} \times (1-x) + v_{X-O(i)} \times x$, ($i=1..3$)

The $v_{X-O1}$, $v_{X-O2}$, $v_{X-O3}$ are calculated as for $v_{A-O1}$, $v_{A-O2}$, $v_{A-O3}$ but with the $R_0$ for bond $X^{2+}-O$. Since the replacing of A by X splits B into 2 different oxidation states 3+ and 4+, we must also replace $v_{B-O1}$, $v_{B-O2}$, $v_{B-O3}$ by their averages:

$<v_{B-O(i)}> = v_{B+3-O(i)} \times (1-x) + v_{B+4-O(i)} \times x$, ($i=1..3$)

**Table 1.** *Parameter $R_0$ for the studied compounds*

| Bond | $R_0$ | Bond | $R_0$ |
|---|---|---|---|
| $La^{3+}-O$ | 2.172 | $Nd^{3+}-O$ | 2.105 |
| $Sr^{2+}-O$ | 2.118 | $Fe^{3+}-O$ | 1.759 |
| $Co^{3+}-O$ | 1.670 | $Na^{1+}-O$ | 1.803 |
| $Co^{4+}-O$ | 1.640 | $Eu^{3+}-O$ | 2.074 |
| $Mn^{3+}-O$ | 1.760 | $Ti^{4+}-O$ | 1.815 |
| $Mn^{4+}-O$ | 1.753 | $Pb^{2+}-O$ | 2.112 |

For the parameters $R_0$ in the above relations, **Table 1** summarizes all used values. They are taken from [4], except for $Co^{3+}$ and $Co^{4+}$, whose values are empirically chosen by the author. The parameter B is set fixed at 0.370 for all cases. The above discussed formalism is limitted to the substitution of the element $X^{2+}$ for $A^{3+}$ in A-positions (0,0,0). The possibility of $X^{2+}$ replacing $B^{3+}$ at (½,½,½), or even replacing $O_1$, $O_2$, $O_3$ at (½,½,0), (½,0,½) and (0,½,½) is not considered.

The present procedure has been applied for all $La_{1-x}Sr_xCoO_3$ samples ($x$=0.0-0.5), for $La_{1-x}Pb_xMnO_3$ with $x$=0.4-0.5 and for several other examples [12, 13, 14, 32] where the precise Rietveld refinement of atomic positions is not available. It requires for these cases the proper transformation of lattice parameters to the pseudo-cubic ones.

## 3 The structural data

### 3.1. Reviewing the pseudo-cubic lattices

(*a*) For $La_{1-x}Pb_xMnO_3$ ($x$=0.1-0.5) the refinement was successful in the pseudo-cubic lattices with the symmetry decreased from cubic ($x$=0.5) to rhombohedral ($x$=0.4) and triclinic ($x$=0.3, 0.2, 0.1). The results are summarized in **Table 2** with the standard deviations given in the parenthesis. For $x$= 0.4-0.5 all atomic positions were fixed; for $x$=0.1-0.3 the metals positions were fixed while the oxygens were refined. The existence of a cubic cell for $x$=0.5 is very similar to of compounds $La_{1-x}Sr_xCoO_3$ [17,18]. The variation of unit cell volumes was $\Delta V/V$=1.0% i.e. max-min variation=0.6Å³. The largest volume occurred for $x$=0.2, 0.3 and the smallest for $x$=0.5. The diversity in lattice constants is 0.5% (from 3.877 to 3.895Å). **Table 3** shows bond lengths and bond angles for Mn–O and O–Mn–O. The average Mn–O–Mn angle increases with $x$ and reaches maximum for $x$=0.4 and *0.5* whereas the length Mn–O decreases continuously to the minimum for *$x$=0.5*.

**Table 2.** *Atomic postions, B and s.o.f. for $La_{1-x}Pb_xMnO_3$*

| $x$ | Atom | $x$ | $y$ | $z$ | $B$ | s.o.f. | S.G. a,b,c α,β,γ | $R_I$ $R_P$ $V$ |
|---|---|---|---|---|---|---|---|---|
| 0.1 | Pb | 0.0 | 0.0 | 0.0 | 0.24 | 0.11 | P1 | 6.2 |
| | La | 0.0 | 0.0 | 0.0 | 0.29 | 0.89 | 3.880(2) | 11.6 |
| | $Mn^{3+}$ | 0.5 | 0.5 | 0.5 | 0.39 | 0.73 | 3.882(1) | 58.6 |
| | $Mn^{4+}$ | 0.5 | 0.5 | 0.5 | 0.39 | 0.27 | 3.890(1) | |
| | O1 | 0.431(2) | 0.500(1) | 0.000 | 1.40 | 0.98 | 90.56(3) | |
| | O2 | 0.500(2) | 0.000 | 0.406(8) | 1.40 | 0.98 | 90.37(3) | |
| | O3 | 0.000 | 0.422(1) | 0.500(4) | 1.40 | 0.98 | 90.54(1) | |
| 0.2 | Pb | 0.0 | 0.0 | 0.0 | 0.24 | 0.21 | P1 | 4.9 |
| | La | 0.0 | 0.0 | 0.0 | 0.29 | 0.81 | 3.887(5) | 10.5 |
| | $Mn^{3+}$ | 0.5 | 0.5 | 0.5 | 0.29 | 0.76 | 3.894(1) | 58.9 |
| | $Mn^{4+}$ | 0.5 | 0.5 | 0.5 | 0.29 | 0.20 | 3.893(1) | |
| | O1 | 0.545(1) | 0.416(4) | -0.001(4) | 1.45 | 1.00 | 90.38(2) | |
| | O2 | 0.481(3) | 0.001(7) | 0.418(5) | 1.45 | 1.00 | 90.42(3) | |
| | O3 | 0.000 | 0.474(4) | 0.450(1) | 1.45 | 1.00 | 90.41(5) | |
| 0.3 | Pb | 0.0 | 0.0 | 0.0 | 0.40 | 0.29 | P1 | 2.8 |
| | La | 0.0 | 0.0 | 0.0 | 0.33 | 0.69 | 3.895(4) | 10.2 |
| | $Mn^{3+}$ | 0.5 | 0.5 | 0.5 | 0.36 | 0.77 | 3.891(1) | 58.9 |
| | $Mn^{4+}$ | 0.5 | 0.5 | 0.5 | 0.36 | 0.21 | 3.882(1) | |
| | O1 | 0.424(6) | 0.493(1) | 0.000 | 1.35 | 0.97 | 90.28(1) | |
| | O2 | 0.568(1) | .000 | 0.507(7) | 1.35 | 0.97 | 90.40(2) | |
| | O3 | -0.004(7) | 0.469(7) | 0.533(5) | 1.35 | 0.97 | 90.37(1) | |
| 0.4 | Pb | 0.0 | 0.0 | 0.0 | 0.40 | 0.38 | Rhombo | 2.5 |
| | La | 0.0 | 0.0 | 0.0 | 0.53 | 0.64 | R $\bar{3}$ c | 9.3 |
| | $Mn^{3+}$ | 0.5 | 0.5 | 0.5 | 0.71 | 0.78 | 3.885(1) | 58.64 |
| | $Mn^{4+}$ | 0.5 | 0.5 | 0.5 | 0.86 | 0.20 | 90.05(5) | |
| | O1 | 0.5 | 0.5 | 0.0 | 1.06 | 1.00 | | |
| | O2 | 0.5 | 0.0 | 0.5 | 1.06 | 1.00 | | |
| | O3 | 0.0 | 0.5 | 0.5 | 1.06 | 1.00 | | |
| 0.5 | Pb | 0.0 | 0.0 | 0.0 | 0.40 | 0.48 | Cubic | 2.3 |
| | La | 0.0 | 0.0 | 0.0 | 0.53 | 0.53 | Pm3m | 8.4 |
| | $Mn^{3+}$ | 0.5 | 0.5 | 0.5 | 0.71 | 0.79 | 3.877(4) | 58.3 |
| | $Mn^{4+}$ | 0.5 | 0.5 | 0.5 | 0.86 | 0.18 | 90 | |
| | O1 | 0.5 | 0.5 | 0.0 | 1.06 | 3.01 | | |

The $La_{1-x}Pb_xMnO_3$ compounds, known to be the feromagnetic materials, have been investigated in [20,21]. Recent interest on these compounds [19,22] was due mainly to the colossal magnetoresistance effect (CMR) discovered in the $Ln_{1-x}A_xMnO_3$ ($Ln^{3+}$= trivalent rare-earth, A= divalent metals $Ca^{+2}$, $Sr^{2+}$, $Ba^{2+}$, $Pb^{2+}$) (see e.g. [23]). In [22] the Rietveld analysis was done for all $x$=0.0-0.5 in the hexagonal space group R $\bar{3}$ c. To compare these results with ours, the pseudo-cubic lattices should be transformed by the transformation matrix [(1,−1,0) (0,1,−1) (2,2,2)]. Except for $x$=0.1 and 0.2 where small angular deformations were seen by our samples (<0.2$^0$) all other cases differed only in the cell constants. By average their cells are 0.5% smaller i.e. 0.3Å$^3$/unit cell. The increment of Mn–O–Mn angles due to substitution was also lesser than in our samples.

**Table. 3** *Bond lengths [Å] and bond angles [$^0$] for $La_{1-x}Pb_xMnO_3$*

| $x$ | to | Mn–O | Mn–O–Mn | $x$ | Mn–O | Mn–O–Mn |
|---|---|---|---|---|---|---|
| 0.1 | O$_1$ | 1.962, 1.965 | 164.3 | 0.3 | 1.964, 1.971 | 162.6 |
|  | O$_2$ | 1.972, 1.979 | 158.7 |  | 1.960, 1.966 | 164.5 |
|  | O$_3$ | 1.961, 1.966 | 162.3 |  | 1.937, 1.974 | 169.7 |
| 0.2 | O$_1$ | 1.979, 1.984 | 158.5 | 0.4 | 1.943 | 180.0 |
|  | O$_2$ | 1.967, 1.979 | 161.0 |  |  |  |
|  | O$_3$ | 1.956, 1.956 | 166.9 | 0.5 | 1.939 | 180.0 |

(*b*) $La_{0.6}Sr_{0.4-x}Ti_xMnO_3$. These structures are very common to of doped lanthanum manganates, although the formula stoichiometry does not argue for such conclusion. It seems that mixing of raw materials according to the given stoichiometry would questionably lead to the replacement of (La, Sr) by Ti$^{4+}$ since the ionic radius of Ti$^{4+}$ (0.42Å, original TiO$_2$) is much smaller than of La$^{3+}$ and Sr$^{2+}$ (1.032Å, 1.18Å) and is comparable to of Mn$^{3+}$, Mn$^{4+}$ (0.58Å, 0.39Å) [26]. The Ti$^{4+}$ fit naturally better into the (Mn$^{3+}$, Mn$^{4+}$) positions. This non-standard mixing of stoichiometry was made intentionally for the study of magnetic properties and one might even doubt whether these compounds were still perovskites. However, the X-Ray diffraction patterns (**Fig. 3**) showed the common peak positions for perovskites, revealing directly the lengths of possible crystallographic axis. All structures were refinable in the rhombohedral space group R$\bar{3}$c with the atomic positions set fixed. **Table 4** lists the results. The replacement of Mn$^{4+}$/Mn$^{3+}$ by the Ti$^{4+}$ was not observed (zeroed s.o.f.). The occurence of Ti$^{2+}$ and/or Ti$^{3+}$ was not also seen.

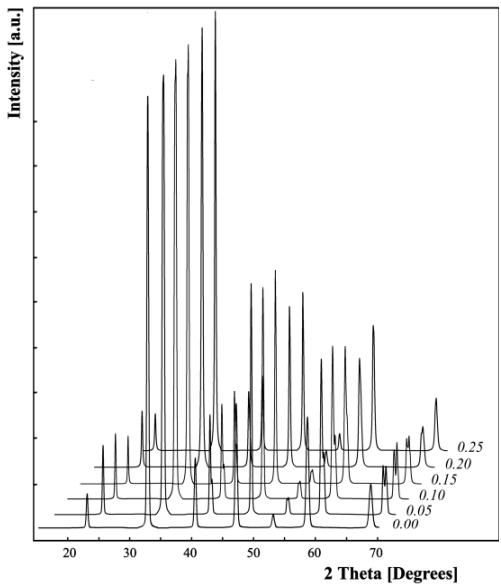

**Fig. 3**. *X-Ray diffraction patterns of $La_{0.6}Sr_{0.4-x}Ti_xMnO_3$.*

The max-min diversity of lattice constants is 0.7%, i.e. 0.03Å and of volumes is 2%, i.e. 1.3Å$^3$. The mean volume <V>=59.2Å$^3$. The average (La, Sr, Ti)–O distances, as deduced from the lattice constants are: 2.744, 2.750, 2.759, 2.764, 2.761, 2.758Å for $x$ =0.0–0.25 sequentially. Similarly, the bond lengths Mn−O are: 1.940, 1.945, 1.951, 1.954, 1.953, 1.950Å. The maximal values occur for $x$=0.15. The bond angles Mn−O−Mn holds fixed at 180$^o$ for all samples.

**Table 4.** *Thermal parameter B and s.o.f. for $La_{0.6}Sr_{0.4-x}Ti_xMnO_3$*

| $x$ | Atom | B | s.o.f. | S.G. $a, \alpha, V$ $R_I, R_P$ | $x$ | B | s.o.f. | S.G. $a, \alpha, V$ $R_I, R_P$ |
|---|---|---|---|---|---|---|---|---|
| 0.00 | La | 0.41 | 0.60 | Rhombo R$\bar{3}$c 3.880(3) 90.1(2) 58.4 3.9 12.7 | 0.15 | 0.30 | 0.59 | Rhombo R$\bar{3}$c 3.908(1) 90.3(1) 59.7 5.2 11.4 |
|  | Sr | 0.77 | 0.40 |  |  | 0.80 | 0.25 |  |
|  | Ti | 0.00 | 0.00 |  |  | 0.53 | 0.15 |  |
|  | Mn$^{3+}$ | 0.30 | 0.72 |  |  | 0.22 | 0.55 |  |
|  | Mn$^{4+}$ | 0.30 | 0.32 |  |  | 0.22 | 0.45 |  |
|  | O1 | 1.22 | 0.97 |  |  | 2.38 | 1.01 |  |
|  | O2 | 1.22 | 0.97 |  |  | 2.38 | 1.01 |  |
|  | O3 | 1.22 | 0.97 |  |  | 2.38 | 1.01 |  |
| 0.05 | La | 0.51 | 0.62 | Rhombo R$\bar{3}$c 3.889(2) 90.3(3) 58.8 6.4 11.7 | 0.20 | 0.57 | 0.61 | Rhombo R$\bar{3}$c 3.905(1) 90.2(2) 59.5 4.8 11.9 |
|  | Sr | 0.85 | 0.35 |  |  | 0.86 | 0.19 |  |
|  | Ti | 0.57 | 0.05 |  |  | 0.60 | 0.19 |  |
|  | Mn$^{3+}$ | 0.30 | 0.71 |  |  | 0.48 | 0.57 |  |
|  | Mn$^{4+}$ | 0.30 | 0.31 |  |  | 0.48 | 0.43 |  |
|  | O1 | 1.34 | 1.00 |  |  | 1.42 | 0.98 |  |
|  | O2 | 1.34 | 1.00 |  |  | 1.42 | 0.98 |  |
|  | O3 | 1.34 | 1.00 |  |  | 1.42 | 0.98 |  |
| 0.10 | La | 0.49 | 0.63 | Rhombo R$\bar{3}$c 3.902(5) 90.3(5) 59.4 5.9 12.9 | 0.25 | 0.27 | 0.60 | Rhombo R$\bar{3}$c 3.900(3) 90.1(1) 59.3 6.8 13.8 |
|  | Sr | 0.31 | 0.30 |  |  | 0.54 | 0.16 |  |
|  | Ti | 0.23 | 0.10 |  |  | 0.59 | 0.27 |  |
|  | Mn$^{3+}$ | 0.29 | 0.61 |  |  | 0.38 | 0.53 |  |
|  | Mn$^{4+}$ | 0.17 | 0.38 |  |  | 0.38 | 0.47 |  |
|  | O1 | 1.72 | 0.98 |  |  | 1.63 | 1.00 |  |
|  | O2 | 1.72 | 0.98 |  |  | 1.63 | 1.00 |  |
|  | O3 | 1.72 | 0.98 |  |  | 1.63 | 1.00 |  |

(*c*) $La_{1-x}Sr_xCoO_3$. The Rietveld analysis was not done for this class of compounds. Here the pseudo-cubic lattices were obtained by transformation of various known structures. For cases where reflection lists were available [15,16] we have re-calculated the lattice parameters and refined them in the pseudo-cubic form. Some differences occured, mostly for the structures published earlier e.g. $La_{0.6}Sr_{0.4}CoO_3$ [15], $La_{0.5}Sr_{0.5}CoO_3$, $La_{0.9}Sr_{0.1}CoO_3$ [16]. Commonly, the transformed lattices agree well with each other (max-min diversity of lattice constants is less than 0.4%, i.e. <0.01Å). A general conclusion can be made as followed. For $x$ =0.0-0.5 the step-by-step changing of the lattice constants from 3.830Å to 3.836Å was observed but was not significant.

**Table 5.** *The pseudo-cubic lattices of $La_{1-x}Sr_xCoO_3$*

| $x$ | Cell | $a$ [Å], $\alpha$ [°] | $V$ [Å$^3$] | Ref. |
|---|---|---|---|---|
| 0.00 | Rhombo | 3.826(2), 90.7(1) | 56.0(3) | [17] |
| 0.10 | Rhombo | 3.832(3), 90.5(3) | 56.3(2) | [15,17] |
| 0.20 | Rhombo | 3.836(2), 90.4(1) | 56.5(1) | [15,17] |
| 0.25 | Orthorh. | 3.831(6), 3.844(8), 3.840(1), 90 | 56.5(3) | [18] |
| 0.30 | Rhombo | 3.834(1), 90.3(1) | 56.3(1) | [16,17] |
| 0.35 | Rhombo | 3.832(3), 90.3(2) | 56.3(1) | [18] |
| 0.40 | Rhombo | 3.831(1), 90.2(2) | 56.2(2) | [15,17] |
| 0.45 | Rhombo | 3.830(4), 90.3(2) | 56.2(2) | [18] |
| 0.50 | Cubic | 3.832(1), 90 | 56.3(2) | [16] |

With respect to those small changes the pseudo-cubic lattices of $La_{1-x}Sr_xCoO_3$ should be regarded as solid, more-less independent to the substitution of $Sr^{2+}$. This is well reflected in the almost constant unit cell volume at all $x$ (mean $<V>=56.3(2)Å^3$). It is worth to notice here that such conclusion is radically different from the common view where the trend is to confirm the influence of substitution on the change of lattice parameters. In solid state physics, the usual practice is to associate phenomena with some phase transition or structural change and the above conclusion is not favourite. Next table summaries results and references.

With respect to the small variations in the lattice constants, the deduced atomic distances vary only a little: from 1.913 to 1.922Å for Co–O and from 2.689 to 2.722Å for (La, Sr)–O. Visibly, there is more change on (La, Sr)–O. The $CoO_6$ coordination spheres can be decribed as regular, rigid and almost non-distorted. The angles Co–O–Co are $180°$ for all samples.

### 3.2. Bond valence and non-stoichiometry

In **Table 6** the atomic bond valences for the given perovskites are shown. Where the atomic positions and interatomic distances are available the BVs were obtained by inserting the bond lengths directly into the formula $v_i=e^{(R_0-R)/B}$. For the other cases, the procedure 2.2. applied. The total bonding electrons in one molecule (i.e. in the asymmetric unit of the pseudo-cubic unit cell) is equal to the total positive charges communicated by all positive cations $\Sigma BE = <v_A> + <v_B>$. Consequently, the expected oxygen stoichiometry is $n(O) = \Sigma BE/2$. The non-stoichiometry parameter $\delta$ is calculated as $\delta = 3 - n(O)$. This $\delta$ directly associates with the valence charge and should correctly be stated as the *charge non-stoichiometry*.

**Table 6.** *Bond valences and non-stoichiometry parameter* $\delta$. *The shading values were calculated using results from Rietveld analysis, other ones were determined from pseudo-cubic lattice constants.*

|  | $x$ | $<v_A>$ | | $<v_B>$ | | $\delta$ | | Meas. $<a>$ | Predic. $a$ $(\delta=0)$ |
|---|---|---|---|---|---|---|---|---|---|
| $La_{1-x}Pb_xMnO_3$ | 0.1 | 2.22 | | 2.44 | 3.63 | 3.42 | 0.08 | 0.14 | 3.879 | 3.863 |
| | 0.2 | 2.23 | | 2.38 | 3.61 | 3.39 | 0.08 | 0.24 | 3.886 | 3.868 |
| | 0.3 | 2.27 | | 2.36 | 3.61 | 3.46 | 0.06 | 0.19 | 3.885 | 3.873 |
| | 0.4 | 2.31 | | 2.38 | 3.62 | 3.63 | 0.03 | -0.02 | 3.885 | 3.878 |
| | 0.5 | 2.38 | | 2.38 | 3.67 | 3.66 | -0.03 | -0.04 | 3.877 | 3.882 |
| $La_{0.6}Sr_{0.4-x}Ti_xMnO_3$ | 0.0 | 2.42 | | 3.66 | | -0.04 | | 3.880 | 3.889 |
| | 0.05 | 2.32 | | 3.62 | | 0.03 | | 3.889 | 3.883 |
| | 0.10 | 2.20 | | 3.57 | | 0.11 | | 3.902 | 3.877 |
| | 0.15 | 2.12 | | 3.55 | | 0.17 | | 3.908 | 3.871 |
| | 0.20 | 2.07 | | 3.57 | | 0.18 | | 3.905 | 3.865 |
| | 0.25 | 2.03 | | 3.58 | | 0.19 | | 3.900 | 3.858 |
| $La_{1-x}Sr_xCoO_3$ | 0.0 | 2.84 | | 3.11 | | 0.02 | | 3.826 | 3.821 |
| | 0.10 | 2.77 | | 3.06 | | 0.08 | | 3.832 | 3.814 |
| | 0.20 | 2.71 | | 3.02 | | 0.13 | | 3.836 | 3.808 |
| | 0.25 | 2.68 | | 3.00 | | 0.16 | | 3.831 | 3.804 |
| | 0.30 | 2.68 | | 3.01 | | 0.16 | | 3.834 | 3.801 |
| | 0.35 | 2.67 | | 3.00 | | 0.16 | | 3.832 | 3.798 |
| | 0.40 | 2.66 | | 2.99 | | 0.17 | | 3.831 | 3.794 |
| | 0.45 | 2.65 | | 2.99 | | 0.18 | | 3.830 | 3.791 |
| | 0.50 | 2.62 | | 2.97 | | 0.21 | | 3.832 | 3.787 |

The $\delta$ measures by itself the *charge deficit* in the unit cell. The predicted cell constants $a$ (assuming rhombohedral case) satisfying condition $\delta=0$, are listed in the last column. These values developed monotonuously with $x$ and differed radically from the measured cell constants. The detailed discussion on this topic is given in Section 4.1. For $La_{1-x}Sr_xCoO_3$ and $La_{0.6}Sr_{0.4-x}Ti_xMnO_3$ the $\delta$ increased with substitution but for $La_{1-x}Pb_xMnO_3$ it decreased. Since for most cases $\delta>0$, the unit cells would contain a bit less valence charge than 6. There are several exceptions with $\delta<0$ near the boundary $x=0.0$ and 0.5. In general the charge deficit could happen by mean of the physical presence of either metals or oxygen vacancies in the lattice. This mechanism, however, always associates with the physical defects and as the charge deficits are not small, the defects themselves must also be compatible. For the single crystals, where the lattice defects are small, this mechanism can hardly lead to the satisfied explanation. An alternative way to understand this charge deficit, avoiding the presence of lattice defects, is to consider that it is principally normal in the real bonding spheres that the atomic valence must not be equal to the atomic stoichiometry.

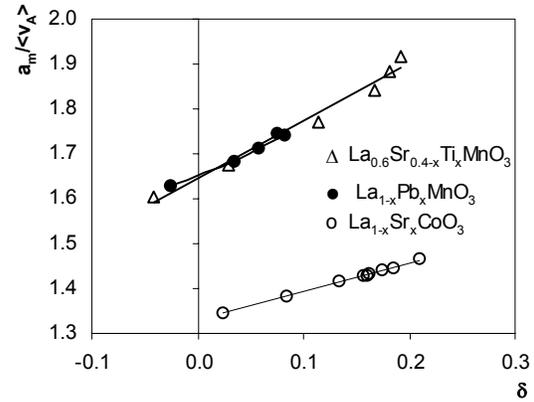

**Fig. 4** *The linear relationship between* $<a>/<v_A>$ *and* $\delta$.

In our samples the substitution seems to have the stronger effect on the change of average valence at A-position $<v_A>$ than at B-position $<v_B>$. The $<v_A>$ varied usually ~ 2 times more then the $<v_B>$, i.e. $\Delta<v_A>/\Delta<v_B>=0.22/0.14$ for $La_{1-x}Sr_xCoO_3$; =0.39/0.09 for $La_{0.6}Sr_xTi_{0.4-x}MnO_3$. The diverged values were found for $La_{1-x}Pb_xMnO_3$: 0.16/0.04 (from lattice) versus 0.06/0.24 (from Rietveld). This forces consideration that the charge flow channel, if existed, must incorporate also the A–$O_{12}$ spheres. **Fig. 4** draws the ratio $<a>/<v_A>$ between the average measured cell constants $<a>$ and the average valences at A-positions $<v_A>$ against the non-stoichiometry parameter $\delta$ for the studied compounds. The dependence is almost linear (correlation coeficient square $R^2>0.98$). For dependence $\delta=f(x)$ (see **Fig. 5**) the observed values did not show clear linear relationship.

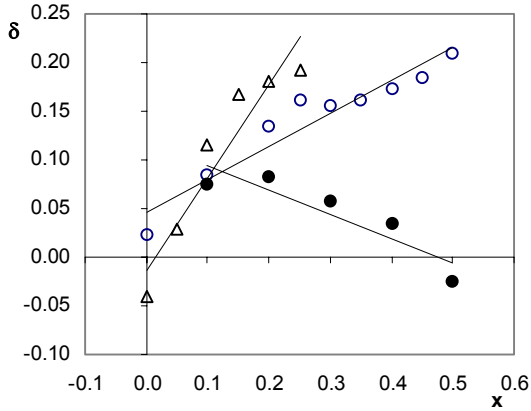

**Fig. 5** *The non-linear dependence of δ on x for the studied perovskites. The legends are same as in Fig.4.*

## 4 Failure of distortion theorem

### 4.1. The development of lattice constants

**Fig. 6** compares the predicted and the measured pseudo-cubic lattice constants for the studied compounds. As distortion theorem predicts, when bond length decreases the number of electrons communicated within the bond must increase. So the lattice tends to compensate unit cell charge deficiency by contracting the lattice constants (inversely, the lattice should expand to reduce charge excess). The measured lattice constants showed, however, no such contraction.

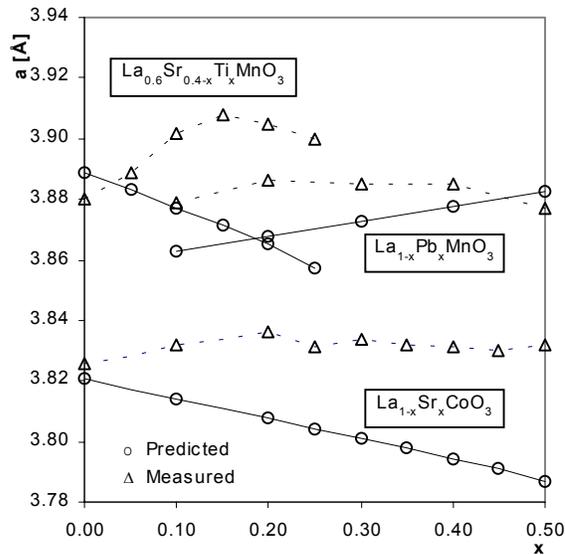

**Fig. 6**. *The predicted lattice constants versus the measured ones. The lines are drawn to guide the eyes.*

The rigidity of the coordination spheres may be expected for $CoO_6$ but not for $MnO_6$ (which is known to be more elastic to Jahn-Teller effect). It seems that the factors which determine the structures are not what drive the valence charges. This incompatibility between the structure and the bonding ability should be seen as the fundamental reason for a non-trivial distribution of the valence charge among the coordination spheres in perovskites.

As seen from Fig. 6, the maximal differences between measured and calculated lattice constants are 0.05Å for $La_{0.6}Sr_{0.4-x}Ti_xMnO_3$, $La_{1-x}Sr_xCoO_3$ and 0.02Å for $La_{1-x}Pb_xMnO_3$. Since the lattice parameters are usually determined with accuracy of third decimal digit, these differences are quite significant. Until they did not decrease below 0.01Å, they should be considered as significant differences. The smallest discrepancy for $La_{1-x}Pb_xMnO_3$ agrees well with a smaller change in the average valence $<v_A>$ and δ for this system. To clasify in terms of the bond valences, the $La_{1-x}Pb_xMnO_3$ can be said to have the least irregular valence charge structure compared to the rest two compound systems. The charge deficit, as represented by δ, is ~0.1 e/u (electron/unit cell) for $La_{1-x}Pb_xMnO_3$ while it is ~0.2 e/u for the rests.

### 4.2. Test for the failure of distortion theorem in other perovskite systems

The validity of distortion theorem was tested for some other perovskite systems. The calculations were performed using the published lattice data and the results are listed in **Table 7**.

**Table 7.** *Maximal difference between the calculated and the measured lattice constant (Δa), maximal valences $<v_A>$, $<v_B>$, maximal δ and the x range for some perovskites*

| Compound | Δa | $<v_A>$ | $<v_B>$ | δ | x range | Ref. |
|---|---|---|---|---|---|---|
| $La_{1-x}Na_xMnO_3$ | 0.04 | 2.12 | 3.72 | 0.04 | 0.1-0.5 | 12 |
| $Ln_{0.8}Sr_{0.2}(Co_{1-x}Fe_x)O_3$ | 0.02 | 2.70 | 3.40 | 0.05 | 0.03-0.12 | 14 |
| $La_{0.5-x}Sr_{0.5}Fe_{0.4}Ti_{0.6}O_{3-δ}$ | 0.07 | 2.62 | 3.64 | 0.07 | 0.0-0.10 | 13 |
| $Nd_{0.67}Sr_{0.33}Mn_{1-x}Fe_xO_3$ | 0.05 | 2.48 | 3.76 | 0.04 | 0.0-0.15 | 31 |
| $CoMnO_{3-δ}$ | 0.08 | 1.96 | 4.46 | -0.17 | 0.0-0.05 | 24 |

For some cases the procedure 2.2 must be modified. Several conclusions can be made for the studied cases: (*a*) for none of the studied samples, the maximal |δ| has decreased below 0.01Å; (*b*) all samples showed the diversity between observed and calculated lattice constants according to the content of substitution *x* but the measure of diversity differed from case to case; (*c*) for most cases, the change of $<v_B>$ was smaller than of $<v_A>$, i.e. the A-positions have located more valence charge than the B-positions.

## 5 Charge distribution

### 5.1. Saturated bond length

There always exists for any oxidation state *v*, calculated as $v = \sum_{i=1}^{n} e^{(R_0-R_i)/B}$ where *i* runs through all *n* bonds in the coordination sphere, the average bond length $R_s$ specifying $v = ne^{(R_0-R_s)/B}$. This $R_s$ is called hereinafter a *saturated bond length* for the given pair $R_0$, *B*. Evidently, $R_s = R_0 - B\log(v/n)$. With respect to the coordination spheres $A-O_{12}$ and $B-O_6$ the saturated bond lengths $R_s(A-O_{12})$ and $R_s(B-O_6)$ set the demarcation line below which the coordination sphere become saturated, i.e. over-charged. Conversely, one should consider the valence charge deficit in the coordination sphere if the larger average bond distances were systematically observed. **Table 8** summaries the saturated bond lengths for several metals frequently encountered in perovskites and **Table 9** compares the measured bond distances with the calculated saturated bond lengths for the studied systems.

**Table 8**. *Some saturated bond lengths for bonding to oxygen*

| Cation | $A-O_{12}$ | $B-O_6$ | Cation | $A-O_{12}$ | $B-O_6$ |
|---|---|---|---|---|---|
| $Ba^{2+}$ | 2.948 | 2.691 | $Cu^{1+}$ | 2.519 | 2.263 |
| $Sr^{2+}$ | 2.781 | 2.524 | $Cu^{3+}$ | 2.252 | 1.995 |
| $Pb^{2+}$ | 2.775 | 2.518 | $Mn^{2+}$ | 2.453 | 2.196 |
| $Ag^{1+}$ | 2.761 | 2.505 | $Mn^{3+}$ | 2.273 | 2.016 |
| $Na^{1+}$ | 2.722 | 2.466 | $Mn^{4+}$ | 2.159 | 1.903 |
| $La^{3+}$ | 2.685 | 2.428 | $Co^{3+}$ | 2.183 | 1.926 |
| $Pr^{3+}$ | 2.651 | 2.394 | $Co^{4+}$ | 2.046 | 1.790 |
| $Hg^{2+}$ | 2.635 | 2.378 | $Pb^{4+}$ | 2.448 | 2.192 |
| $Ca^{2+}$ | 2.630 | 2.373 | $Ti^{4+}$ | 2.221 | 1.965 |
| $Nd^{3+}$ | 2.618 | 2.361 | $Fe^{2+}$ | 2.397 | 2.140 |
| $K^{1+}$ | 3.051 | 2.795 | $Cr^{6+}$ | 2.050 | 1.794 |
| $Er^{3+}$ | 2.501 | 2.244 | $Fe^{3+}$ | 2.272 | 2.015 |
| $Eu^{3+}$ | 2.587 | 2.330 | $Fe^{4+}$ | 2.186 | 1.930 |

According to Table 8, the cations listed in the left part of the table can hardly occupy B-positions since their saturated bong lengths are much larger than the average bond lengths observed for the $B-O_6$ spheres. If this would happen, these spheres would become heavily saturated. Similarly, if the cations listed in the right part of the table occur in $A-O_{12}$ spheres then the resulting coordinations must be much non-saturated because the listed saturated bond lengths are significantly smaller than the observed average bond lengths for these spheres.

**Table 9**. *The average bond lengths and the saturated ones in the spheres $A-O_{12}$ and $B-O_6$ for the studied perovskites*

| | x | $A-O_{12}$ | | | $B-O_6$ | | |
|---|---|---|---|---|---|---|---|
| | | Meas. | Satur. | $\Delta_{M-S}$ | Meas. | Satur. | $\Delta_{M-S}$ |
| $La_{1-x}Pb_xMnO_3$ | 0.1 | 2.756 | 2.694 | 0.062 | 1.967 | 2.005 | -0.038 |
| | 0.2 | 2.760 | 2.703 | 0.057 | 1.970 | 1.994 | -0.024 |
| | 0.3 | 2.757 | 2.712 | 0.045 | 1.962 | 1.982 | -0.020 |
| | 0.4 | 2.747 | 2.721 | 0.026 | 1.943 | 1.971 | -0.028 |
| | 0.5 | 2.742 | 2.730 | 0.012 | 1.939 | 1.960 | -0.021 |
| $La_{0.6}Sr_{0.4-x}Ti_xMnO_3$ | 0.00 | 2.744 | 2.723 | 0.021 | 1.940 | 1.971 | -0.031 |
| | 0.05 | 2.750 | 2.695 | 0.055 | 1.945 | 1.982 | -0.037 |
| | 0.10 | 2.759 | 2.667 | 0.092 | 1.951 | 1.994 | -0.043 |
| | 0.15 | 2.763 | 2.639 | 0.124 | 1.954 | 2.005 | -0.051 |
| | 0.20 | 2.761 | 2.611 | 0.150 | 1.952 | 2.016 | -0.064 |
| | 0.25 | 2.757 | 2.583 | 0.174 | 1.950 | 2.028 | -0.078 |
| $La_{1-x}Sr_xCoO_3$ | 0.00 | 2.706 | 2.685 | 0.021 | 1.913 | 1.926 | -0.013 |
| | 0.10 | 2.710 | 2.695 | 0.015 | 1.916 | 1.913 | 0.003 |
| | 0.20 | 2.713 | 2.704 | 0.009 | 1.918 | 1.899 | 0.019 |
| | 0.25 | 2.714 | 2.709 | 0.005 | 1.919 | 1.892 | 0.027 |
| | 0.30 | 2.711 | 2.714 | -0.003 | 1.917 | 1.886 | 0.031 |
| | 0.35 | 2.710 | 2.719 | -0.009 | 1.916 | 1.879 | 0.037 |
| | 0.40 | 2.709 | 2.723 | -0.014 | 1.916 | 1.872 | 0.044 |
| | 0.45 | 2.708 | 2.728 | -0.020 | 1.915 | 1.865 | 0.050 |
| | 0.50 | 2.710 | 2.733 | -0.023 | 1.916 | 1.858 | 0.058 |

The data given in Table 9 reveal that the valence charge does not locate symmetrically on $A-O_{12}$ and $B-O_6$. Particularly, for $La_{0.6}Sr_{0.4-x}Ti_xMnO_3$ and $La_{1-x}Pb_xMnO_3$ the $Mn-O_6$ spheres are over-charged, whereas the $(La,Sr,Ti)-O_{12}$ and $(La,Pb)-O_{12}$ spheres are under-charged. The situation is inverted for $La_{1-x}Sr_xCoO_3$: the $Co-O_6$ spheres are under-charged and the $(La, Sr)-O_{12}$ spheres exhibit both under-charged and over-charged state.

For all cases, one can observe that when the spheres $B-O_6$ develop closer to saturation level ($\Delta_{M-S} \to 0$), the $A-O_{12}$ also come closer to this level, that is to say, when some charge flows away from $B-O_6$ spheres, the $A-O_{12}$ also receive some more charge and *vice-versa*. Evidently, this shift minimalizes the overall decline from saturation of bonds forced by substitution. Note that the bonding saturation of $B-O_6$ and $A-O_{12}$ does not develop in parallel with the average valences of these spheres. The average valence may decrease (see Table 6, $<v_A>$ of $La_{1-x}Sr_xCoO_3$) while the sphere is going more saturated (see Table 9, $A-O_{12}$ of $La_{1-x}Sr_xCoO_3$ or Fig.7, $(La,Sr)-O_{12}$).

### 5.2. Saturated bond valence

The atomic valence associated with the saturated bond length $R_s$ is called the saturated bond valence, or for short the saturated valence $v_s$. These saturated valences can be calculated from the data listed in Table 9 according to the formula $v_s = v/n = e^{(R_0-R_s)/B}$. The results are collected in **Table 10**.

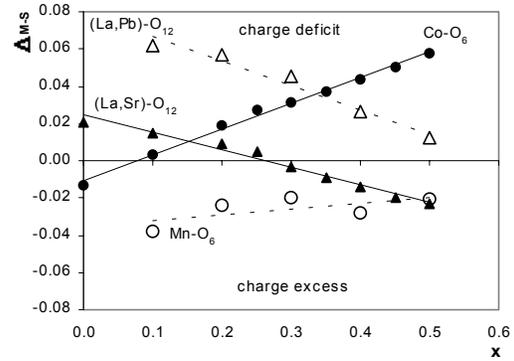

**Fig. 7**. *The linear dependences of $\Delta_{M-S}$ (difference between the measured and the saturated bond lengths) on the content of substitution x for $La_{1-x}Pb_xMnO_3$ and $La_{1-x}Sr_xCoO_3$. The draw for $La_{0.6}Sr_{0.4-x}Ti_xMnO_3$ is omitted for clarity since it is similar to of $La_{1-x}Pb_xMnO_3$.*

The difference between measured and saturated valence is expressed as $\Delta v_{A-S} = <v_A> - v_s(A-O_{12})$ and $\Delta v_{B-S} = <v_B> - v_s(B-O_6)$ respectively. The total valence deviation $\Delta v$ of both spheres, calculated as $\Delta v = \Delta v_{A-S} - \Delta v_{B-S}$ is drawn against the content of substitution $x$ in **Fig. 8**. This $\Delta v$ illustrates the balancing of the valence charge between the spheres $B-O_6$ and $A-O_{12}$. As $\Delta v$ increases, the charge is shifted from $B-O_6$ to $A-O_{12}$ and *vice-versa*.

For $La_{1-x}Sr_xCoO_3$ the graph crosses $x$ axis near $x \sim 0.15$. This means that at this substitution the coordination spheres of metals decline equally from the saturation, whereas for other cases the spheres are differently saturated. The measure of this difference is expressed in the unit of electrons per unit cell and explores the asymmetric distribution of the valence charge between the coordination spheres. The change of this measure describes the amount of charge flowed between them.

**Table 10**. *The deviation of the measured average bond valences from the saturated valences for the studied systems*

| | $x$ | $A$–$O_{12}$ | | $B$–$O_6$ | | $\Delta v$ |
|---|---|---|---|---|---|---|
| | | $v_s(A$–$O_{12})$ | $\Delta v_{A-S}$ | $v_s(B$–$O_6)$ | $\Delta v_{B-S}$ | |
| $La_{1-x}Pb_xMnO_3$ | 0.1 | 2.88 | -0.44 | 3.09 | 0.33 | -0.78 |
| | 0.2 | 2.77 | -0.40 | 3.18 | 0.21 | -0.61 |
| | 0.3 | 2.66 | -0.30 | 3.27 | 0.18 | -0.49 |
| | 0.4 | 2.56 | -0.17 | 3.37 | 0.26 | -0.44 |
| | 0.5 | 2.46 | -0.08 | 3.46 | 0.20 | -0.28 |
| $La_{0.6}Sr_{0.4-x}Ti_xMnO_3$ | 0.00 | 2.56 | -0.14 | 3.37 | 0.29 | -0.44 |
| | 0.05 | 2.69 | -0.37 | 3.27 | 0.34 | -0.72 |
| | 0.10 | 2.83 | -0.62 | 3.18 | 0.39 | -1.01 |
| | 0.15 | 2.97 | -0.84 | 3.09 | 0.46 | -1.30 |
| | 0.20 | 3.11 | -1.04 | 3.00 | 0.57 | -1.60 |
| | 0.25 | 3.26 | -1.22 | 2.91 | 0.68 | -1.91 |
| $La_{1-x}Sr_xCoO_3$ | 0.00 | 3.00 | -0.17 | 3.00 | 0.11 | -0.27 |
| | 0.10 | 2.88 | -0.11 | 3.09 | -0.02 | -0.09 |
| | 0.20 | 2.77 | -0.07 | 3.18 | -0.16 | 0.09 |
| | 0.25 | 2.72 | -0.04 | 3.23 | -0.23 | 0.19 |
| | 0.30 | 2.66 | 0.02 | 3.27 | -0.26 | 0.28 |
| | 0.35 | 2.61 | 0.06 | 3.32 | -0.32 | 0.38 |
| | 0.40 | 2.56 | 0.10 | 3.37 | -0.38 | 0.48 |
| | 0.45 | 2.51 | 0.14 | 3.42 | -0.43 | 0.57 |
| | 0.50 | 2.46 | 0.16 | 3.47 | -0.50 | 0.66 |

With content of $Sr^{2+}$ increased from 0 to 0.5 each sphere $(La,Sr)$–$O_{12}$ in $La_{1-x}Sr_xCoO_3$ gained $|0.16 - (-0.17)| \sim 0.3e^-$ whereas the sphere $Co$–$O_6$ lost $|-0.50 - 0.11| \sim 0.6e^-$. At least $0.3e^-$ should be considered as shifted from $Co$–$O_6$ to $(La,Sr)$–$O_{12}$ although twice such amount has moved away from $Co$–$O_6$. Here the substitution has mainly effected the $Co$–$O_6$ spheres. For $La_{1-x}Pb_xMnO_3$ the $(La,Pb)$–$O_{12}$ gained $|-0.08-(-0.44)| \sim 0.4e^-$ and the $Mn$–$O_6$ lost $|0.20-0.33| \sim 0.1e^-$ when content of $Pb^{2+}$ increased. The shifted amount of charge should be $0.1e^-$. Here the substitution had more influence on $(La,Pb)$–$O_{12}$ spheres than on the $Mn$–$O_6$. In $La_{0.6}Sr_{0.4-x}Ti_xMnO_3$ the spheres behaved similarly as in $La_{1-x}Pb_xMnO_3$ if the substitution was taken in reverse order, i.e. from 0.25 downto 0.0. The charge deviation and charge shift were however nearly 3-4 times larger: $|-1.22-(-0.14)| \sim 1.1e^-$ for $(La,Sr,Ti)$–$O_{12}$ and $|0.68 -0.29| \sim 0.4e^-$ for $Mn$–$O_6$. Again, the $A$–$O_{12}$ spheres have located more valence charge.

While both spheres in $La_{1-x}Pb_xMnO_3$ and $La_{0.6}Sr_{0.4-x}Ti_xMnO_3$ become closer to the saturation balance, the spheres in $La_{1-x}Sr_xCoO_3$ developed away from this level. The maximal total valence deviation $\Delta v$ was near $0.8e^-$ for $La_{1-x}Pb_xMnO_3$, $0.7e^-$ for $La_{1-x}Sr_xCoO_3$ and $\sim 2e^-$ for $La_{0.6}Sr_{0.4-x}Ti_xMnO_3$.

The valence deviation from saturation of the spheres must be seen as the intrinsic bonding property of perovskites. It forces the consideration that, by its nature the valence charge is polarized, although no such argumentation is found in its definition. In all studied samples, the metals behave more-less like oppositely charged ions: the La, Pb, Sr, Ti are positive and the Co, Mn are negative. Since the valence charge are averaged charge, the real local charge asymmetries of the spheres can certainly deviate more than $2e^-/u$ from the saturation balance. Thus some charge flow channel may be created by these local asymmetries.

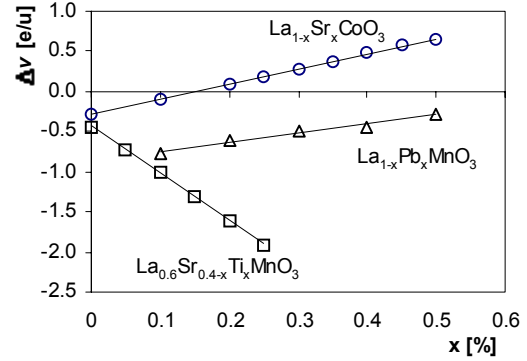

**Fig. 8**. *The strong linear dependences of $\Delta v$ on substitution x show the balancing of valence charge in the spheres. The $\Delta v$ is expressed in the unit of electrons per unit cell. The linearity also exists between the $\Delta v_{B-S}$, $\Delta v_{A-S}$ and x but we omit them here for clarity.*

## 6 Problems and remarks

### 6.1. Resistibility to noise and accuracy of method

Since the determination of the coeficients $R_0$ was done statistically, the bond valence method is expected to have good resistibility to noise. The possible disturbance of results by the systematic or random errors (during samples preparation, measurement and structure determination) is believed to be limited by the averaging of inputs when the $R_0$-s were determined. As tested and reported in [5], the bond valence method has accuracy around 5-7% for the ionic compounds. We have tested this method against the complex compounds (covalence bonding character prevails) and found its accuracy was even a little better at 5-6%. Evidently, this accuracy does not depend on the physical meaning of bond valence but only on the statistical rules applying to the coeficients $R_0$ and B, i.e. on the process of averaging input data. Note that for perovskites the frequently mentioned Rietveld method provides $R_{Profile} \sim 10\%$. However, the bond valence method has one limitation just because of its statistical origin, that is it can not be applied everywhere *a priori*. The parameters obtained using one statistical file must not imply the correct results for the other files. This method is risky when using to analyse one single case but is powerful for studying the collective effect of many samples.

### 6.2. The charge non-stoichiometry δ versus the real non-stoichiometry x

It is not clear how can the charge non-stoichiometry δ be associated with the real structure defects, although the occurences of defects certainly affect δ. For the cases studied above, δ does not imply any real defects. In principle, charge non-stoichiometry can happen in the perfect structures without any defects. The next analysis can illustrate.

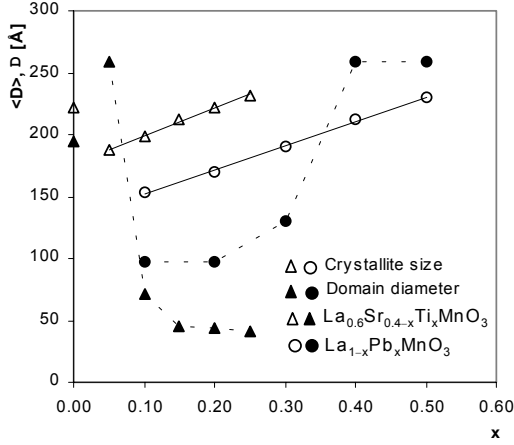

**Fig. 9**. *The crystallite sizes as determined from X-Ray study and the domain diameter* D.

By suggestion that the charge deficit was generated by the oxygen holes in the lattice, the holes density ρ could be directly determined from δ. For example, δ = 0.16 $e^-$/u for $La_{0.7}Sr_{0.3}CoO_3$ (Tab. 6) means that every $n$=12.5 unit cells produces 1 missing oxygen (since the oxidation state of the oxygens is 2− and 2/0.16=12.5). So ρ = 1/$n$=0.08 (oxygen/unit cell). Now suggest that the holes are equally distributed inside lattice, so their cross links creat a mosaic network which divides the lattice into the domain areas with the boundaries laid on these links. The maximal diameter of the domains is thus D = $n \times \langle a \rangle = \langle a \rangle / \rho$ = 3.834×12.5 = 47.9Å ($\langle a \rangle$ is the average pseudo-cubic lattice constant). This system would limit the crystallite size and the final effect would be seen by the X-Ray diffraction of the compound. **Fig.9** compares the crystallite size $\langle D \rangle$ determined by the fourier analysis of the strongest diffraction and the domain diameter D for $La_{1-x}Pb_xMnO_3$ and $La_{0.6}Sr_{0.4-x}Ti_xMnO_3$. As seen, for both systems the crystallite size $\langle D \rangle$ follows linearly the content of substitution $x$, whereas the D develops differently. The sharp linear relationship between $\langle D \rangle$ and $x$ for the studied cases is somehow surprising, although some weaker relationships have already been reported (e.g. see [11]). This linearity can hardly be explained by the real defects. It is also not easier for the valence charge.

If we would agree to that a greater crystallite creates a larger flow place for the valence charge transfer, the problem would look more acceptable. This would mean that the crystal growth always minimizes the decline from saturation of spheres and the final crystallite size corresponds to the strongest effect of this minimalization (in given experimental conditions).

### 6.3. Saturated bond valence and $B^{4+}$ concentration

The expected concentration of the cations $B^{4+}$ is equal to the content of substitution $x$ in the studied systems except for $La_{0.6}Sr_{0.4-x}Ti_xMnO_3$ where it is 0.4–$x$. The linear dependences $\Delta v_{B-S}=f(x)$ and $\Delta v_{A-S}=f(x)$ also mean the linearity between $\Delta v_{B-S}$, $\Delta v_{A-S}$ and the expected concentration of $B^{4+}$.

If we now consider $|\Delta v_{B-S}|$ as reflecting the real $B^{4+}$ content, then (*a*) for $La_{1-x}Sr_xCoO_3$ this content correlates well with $x$ but does not exceed $x$ (Table 10, $\Delta v_{B-S} < x$) so the exceeded amount of charge ($\Delta v > x$ for $x$=0.0, 0.35-0.50) comes from the A–$O_{12}$ spheres (0.3$e^-$); (*b*) for the $La_{1-x}Pb_xMnO_3$ and the $La_{0.6}Sr_{0.4-x}Ti_xMnO_3$ this content decreases merely when the stoichiometric concentration $Mn^{4+}$ increases, so the exceeded amount of charge (total $\Delta v > x$) again comes from the A–$O_{12}$ spheres (0.1$e^-$, 1.1$e^-$); (c) the s.o.f.s of $Mn^{4+}$ reported in the Table 2 & Table 4 agree better with the $\Delta v_B$ then with the stoichiometric concentration. Thus we can conclude here that for the studied cases the $\Delta v_B$ is closer to the real $B^{4+}$ concentration then the stoichiometric concentration.

### 6.4. Remark on anisotropy of valence charge

The concepts of charge are commonly divided into two types: the static charge and the dynamic charge. The valence charge corresponds to the static charge. Empirically and in symmetrical bonding spheres this charge can not be polarized. But as has been showed here, it is asymmetrically located, so is polarized in the scale that is not negligible. The valence charge polarization however does not come from any anisotropy in the *metal-oxygen* bonds, as for the dynamic charge, but from anisotropy in the triple bonding system *metal-oxygens-metal*. Note that the word "oxygens" is written in plural.

**Table 11**. *Comparison of the evolution of Born effective charges (1st row) and valence charge (2nd row) for $BaTiO_3$ under isotropic pressure in cubic phase. Cell constant is listed in the first column.*

| $a[Å]$ | $Z_{Ba}$ | $Z_{Ti}$ | $Z_{O\perp}$ | $Z_{O\parallel}$ | $Z(\Sigma_O/2)$ | $Z_{Ba}+Z_{Ti}$ | $\Delta v$ |
|---|---|---|---|---|---|---|---|
| 3.64 | 2.95 | 7.23 | -2.28 | -5.61 | -3.95 | 10.18 | |
|  | 5.49 | 5.92 |  |  | -3.80 | 11.41 | 1.58 |
| 3.94 | 2.77 | 7.25 | -2.15 | -5.71 | -3.93 | 10.02 | |
|  | 3.10 | 3.95 |  |  | -2.35 | 7.04 | 1.15 |
| 4.00 | 2.74 | 7.29 | -2.13 | -5.75 | -3.94 | 10.03 | |
|  | 2.76 | 3.64 |  |  | -2.13 | 6.40 | 1.13 |
| 4.40 | 2.60 | 7.78 | -2.03 | -6.31 | -4.17 | 10.38 | |
|  | 1.29 | 2.12 |  |  | -1.14 | 3.41 | 1.17 |

The Double-Exchange mechanism, frequently used to discuss the magnetic behaviour of perovskites, follows from the simple triple bonding system metal-oxygen-metal (e.g. Mn–O–Mn). The valence charge polarization depends on anisotropy of the whole *chain of bonding spheres*, not of single bonds and is measured by $\Delta v$. **Table 11** compares the sensitivity of the valence charge and the dynamic charge to the structural changes for $BaTiO_3$. The calculation of the dynamic charges come from [27]. The studies on valence charge polarization may lead to explanation of many interesting properties of perovskites.

## 7 Conclusion

We have shown the importance of the pseudo-cubic lattices to perovskites and have determined such structures for two perovskite systems $La_{1-x}Pb_xMnO_3$ ($x$=0.0–0.5) and $La_{0.6}Sr_xTi_{0.4-x}MnO_3$ ($x$=0.0–0.25). The revision of the structures $La_{1-x}Sr_xCoO_3$ ($x$=0.0–0.5) have also been made. The pseudo-cubic lattice is sensitive to the angular deformation and is suitable for the study of small angular deformation in the perovskite lattices. Based on the crystal structure data, a systematic study of valence charge was made together with the theoretical model for retrieving the valence charge information directly from the pseudo-cubic lattices. Several conclusions can be made as followed:

(*a*) The charge non-stoichiometry parameter $\delta$ was always positive and did not reach above 0.02 in the studied systems. No association of this $\delta$ and the real formula stoichiometry $x$ has been observed. This charge non-stoichiometry $\delta$ should occur also in the perfect crystals where the formula would be exactly $ABO_3$.

*b*) All studied perovskites showed the *valence charge deficit* in the unit cell. The clear evidence of this was that there was no visible contraction of the lattice constants in the given systems when the total valence charge in the unit cell increased (i.e. when the non-stoichiometry parameter $\delta \to 0$). This demonstrates the failure of distortion theorem in all these systems.

(*c*) In all studied samples the valence charge was *asymmetrically localized* - with respect to the saturation balance of the bond distances, between the coordination spheres $A-O_{12}$ and $B-O_6$. The dependence of $\Delta_{M-S}$ (the difference between the measured and the saturated bond lengths) on the content of substitution $x$ was clearly linear.

(*d*) As the content of substitution $x$ varied, a certain *charge transfer process* between the coordination spheres $A-O_{12}$ and $B-O_6$ was seen - with respect to the saturation balance of the bond valences. This process is strongly linear to the substitution, as was demonstrated by the linear relationship between the total valence deviation $\Delta v$ and the content of substitution $x$ ($R^2 \sim 1$). The shifted charge was from $0.1e^-$ to $1.1e^-$ per unit cell.

This work involved the theoretical concepts of the saturated bond length and the saturated bond valence. In general they coincide with the statistical averages of the bond lengths and the bond valences of the same bonds. The declines from these average levels signify the state of being under-charged or over-charged of the bonds. In certain aspect these levels mark the bond ability to absorb more electron. With respect to them, a certain internal charge transfer process between the coordination spheres of the metals has been observed. How this process changes under the specific conditions is still the question.

## Acknowledgement

The author would like to thank the Center for Materials Science, Faculty of Physics, HUS-VNU for providing materials and kind helps during measurement and other experimental works. Especial thank is expressed to Prof. Nguyen Chau for fruitful discussions and supports.